# Near-Infrared-Triggered Photodynamic Antibacterial Therapy Using Rose Bengal-Coated Upconverting Nanoparticles

**Pratik Deshmukh[1,2*], Nandini Ahuja[1], Bhumika Sharma[1], Khageswar Sahu[1,2], Srinibas Satapathy[1,2*] and Shovan Kumar Majumder[1,2]**

[1]*Laser Biomedical Applications Division, Raja Ramanna Centre for Advanced Technology, Indore,452013, Madhya Pradesh, India.*

[2]*Homi Bhabha National Institute, Training School Complex, Anushakti Nagar, Mumbai 400094, Maharashtra, India.*

***Address for Correspondence:**

**E-mail Address:** ***ppdeshmukh@rrcat.gov.in***, ***srinu73@rrcat.gov.in***

**Near-Infrared-light Triggered Antimicrobial Photodynamic Therapy Using Rose Bengal-Coated Upconverting Nanoparticles**

**Abstract:**

Antimicrobial photodynamic therapy (aPDT) is a promising modality for inactivation of antibiotic resistant bacteria, relying on the activation of a photosensitizer (PS) by light of a specific wavelength. This process results in the formation of reactive oxygen species, which ultimately induce cell death. However, aPDT in its conventional form, is limited by the shallow penetration of visible light, restricting its effectiveness for treatment of soft tissue and orthopaedic tissues. To overcome this limitation, near-infrared (NIR) absorbing PS can be used. However, poor stability *in vivo* after injection, ineffective microbial targeting due to hydrophilic nature and off-site tissue damage are the issues with use of NIR absorbing bare PSs. This issue can be mitigated by combining NIR light with a upconverting nanoparticles (UCNPs), which mediate in conversion of NIR into visible light for effective PS activation. In this study, $LaF_3:Er^{3+},Yb^{3+}$ nanoparticles were synthesized using a hydrothermal method and coated with Rose Bengal (RB), a promising hydrophilic PS for aPDT, to evaluate the potential for NIR-triggered aPDT. Characterization of the synthesized UCNPs confirmed the crystalline structure, size distribution and successful RB functionalization. Photophysical studies demonstrated efficient energy transfer between UCNPs and RB, leading to singlet oxygen ($^1O_2$) generation in vitro. Antibacterial studies against Methicilin resistant *Staphylococcus aureus (MRSA)*, a superbug of implicated soft tissue and orthopaedic infections, revealed significant photo-bactericidal efficacy upon NIR irradiation, indicating the potential of RB-coated UCNPs for aPDT applications.

## GRAPHICAL ABSTRACT

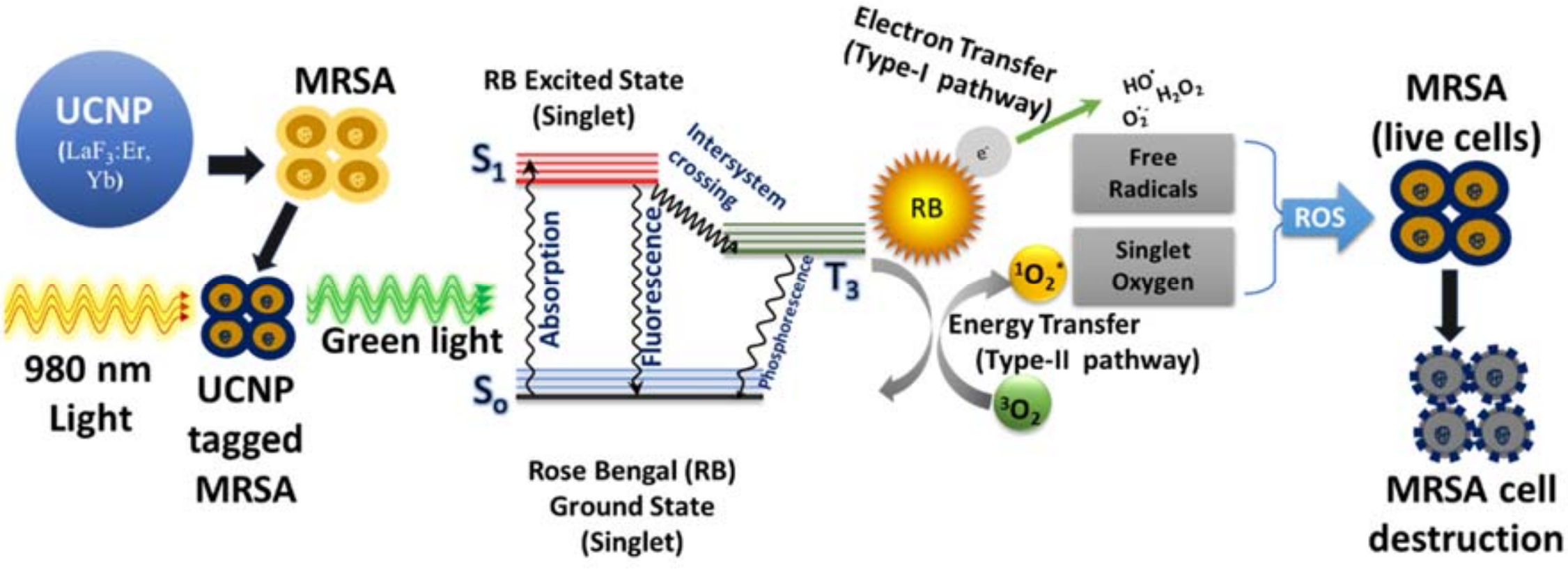

## Introduction

Antimicrobial resistance (AMRs) is now a global scourge and has spurred urgent search of novel antimicrobial strategies against which resistance cannot be generated. Antimicrobial Photodynamic therapy (aPDT) has emerged as a valuable treatment modality for a range of malignancies, owing to its minimally invasive nature, reduced toxicity profile and the potential for repeated applications. [1–3] This therapeutic approach hinges on the selective exposure of diseased target cells or tissues to a photosensitizer (PS), a typically non-toxic dye. Subsequent irradiation with light of a specific wavelength, in the presence of adequate molecular oxygen, triggers a cascade of photochemical reactions.[1,4] Ultimately, this triggers the formation of reactive oxygen species (ROS), the principal agents causing cellular demise within the targeted tissue.[3,5]

The fundamental mechanism of aPDT relies on the light-induced activation of a PS molecule bound to microbes and microbial biofilms.[4,6,7]. Upon illumination with light of an appropriate wavelength, the PS absorbs photons and transitions from its ground state to an excited singlet state (S1). From this excited state, the PS can return to its ground state through fluorescence or internal conversion, or it can undergo intersystem crossing to a longer-lived triplet state (T1, T3). In this triplet state, the PS participates in one of two distinct reaction pathways. A type I reaction involves the formation of radical ions, which subsequently generate ROS such as superoxide and hydroxyl radicals. Alternatively, a type II reaction leads to the direct production of singlet oxygen. These ROS inflict cellular damage by oxidizing critical biomolecules, including membrane lipids, proteins, and nucleic acids, ultimately disrupting cellular function and leading to cell death.[7]

Rose Bengal (RB) is a promising PS for aPDT as it has good triplet yield. However, despite its effectiveness in microbial inactivation, a major limitation is that RB is primarily activated by green light, which restricts its tissue penetration and therapeutic applicability.[8,9] This portion of the electromagnetic spectrum exhibits limited tissue penetration, effectively restricting application of aPDT till date to superficial, intra-epidermal infections/lesions[9]. To address this limitation, the use of near-infrared (NIR) light has been proposed as a promising alternative. NIR light offers superior tissue penetration capabilities and a higher signal-to-noise ratio due to reduced autofluorescence.[10] Directly injecting NIR light-activatable PS, whether hydrophobic or hydrophilic, faces hurdles such as decreased stability, clumping and unintended distribution within the body. These problems hinder the accumulation of sufficient PS at the desired location, thereby limiting the production of ROS. [9,10] As an alternative, attaching the PS to a transducer could allow for its activation using NIR light. This transducer acts as an intermediary, converting the incident NIR light into the visible light required to excite the photosensitizer.[9,11]

The efficient transducers for conversion of NIR to visible are upconversion nanoparticles (UCNPs).[11] As it involve anti-Stokes emission by sequential absorption of two or more incident NIR photons.[12,13] Further, these are excellent luminescent probes because of their unique optical properties such as non-blinking or stable emission, low autofluorescence background, tunable emission, high photostability and deep tissue penetration into biological system.[12,14]. Unlike bare PS, UCNPs offer tunable emission, mitigating autofluorescence interference.[15] Furthermore, UCNPs generally exhibit small size, strong and narrow visible emission, long luminescence lifetimes, high photostability and low cytotoxicity. [16] Conjugating PSs to UCNPs also prevents PS degradation, aggregation and off-target accumulation.[15]

UCNPs are typically composed of lanthanide ions embedded in a nanoscale crystalline inorganic host material, generally ranging from 1 to 100 nm in size.[16] These lanthanide ions exhibit distinct energy level structures that facilitate the anti-Stokes luminescence mechanism.[16,17] When exposed to multiple low-energy photons, they sequentially absorb energy and emit a single photon of higher energy. The efficiency of this upconversion process is heavily dependent on the phonon energy of the host material; lower phonon energies help suppress non-radiative losses, resulting in enhanced luminescence. Among the various host materials, fluoride-based compounds are often preferred for NIR to visible upconversion due to their inherently low phonon energies, which are more favourable than those of oxides or sulphides.[17,18]

Within the lanthanide series, erbium ions ($Er^{3+}$) have been widely investigated as active dopants in phosphors due to their well-defined energy levels. When excited at approximately 970 nm, $Er^{3+}$ ions emit light in both the green ($^2H_{11/2}$ / $^4S_{3/2} \rightarrow {}^4I_{15/2}$) and red ($^4F_{9/2} \rightarrow {}^4I_{15/2}$) regions of the visible spectrum.[19] However, a key limitation of $Er^{3+}$ is its relatively low absorption cross-section, which restricts its efficiency for practical applications.[19,20]

One approach to improve absorption involves increasing the concentration of lanthanide ions within the host material. However, excessive doping can lead to concentration quenching, where non-radiative energy transfer between neighbouring ions results in diminished fluorescence output.[21] To mitigate this issue, researchers often use a combination of lanthanide dopants. In this strategy, a sensitizer ion (such as $Yb^{3+}$ or $Nd^{3+}$) efficiently absorbs NIR light and transfers the energy to an activator ion (such as $Er^{3+}$, $Ho^{3+}$ or $Tb^{3+}$), which then emits the final luminescence.[20,21] This energy transfer process helps optimize the upconversion efficiency while minimizing the negative effects of concentration quenching. Notably, to the best of our knowledge, studies exploring RB-coated Er,Yb co-doped $LaF_3$ nanophosphors as agents for aPDT remain limited, highlighting the novelty and potential of this approach.

In this study, we synthesized RB-coated $Er^{3+}$, $Yb^{3+}$ co-doped $LaF_3$ nanophosphors using a hydrothermal method to ensure uniformity in particle size and morphology. The RB coating was employed to assess the nanophosphor's potential for aPDT applications. The objective of this research is to evaluate the effectiveness of these engineered nanophosphors in facilitating NIR-induced aPDT, thereby advancing biological therapy technologies and expanding the applicability of aPDT to deeper-seated infections.

**Experimental:**

Nanoparticle Synthesis:

$LaF_3$ nanoparticles doped with $Er^{3+}$ and $Yb^{3+}$ ($LaF_3$:Er,Yb) were synthesized using previously established method with slight modifications.[22] These modifications were introduced to enhance control over nanoparticle size and reduce agglomeration.

Lanthanum oxide ($La_2O_3$, 99.99%), erbium oxide ($Er_2O_3$, 99.99%) and ytterbium oxide ($Yb_2O_3$, 99.99%) were procured from Alfa Aesar. Ammonium fluoride ($NH_4F$, 98%) and nitric acid (69%) were sourced from Merck, while ammonia solution ($NH_4OH$, 25%) was obtained from CDH. Rose Bengal (RB, 95%) was acquired from Sigma-Aldrich. All chemicals were utilized without further purification.

The synthesis was initiated by preparing a lanthanide precursor solution, which involved dissolving $Er_2O_3$, $Yb_2O_3$ and $La_2O_3$ in dilute nitric acid (2 M) in stoichiometric proportions (e.g., 2 mol% Er & 20 mol% Yb). A fluoride solution was then prepared by dissolving $NH_4F$ in deionized water and adjusting the pH to 8 using a few drops of $NH_4OH$ solution.
The lanthanide precursor solution was gradually added to the fluoride solution at a controlled rate of 5 ml/min while maintaining the temperature at 50°C. This process resulted in a milky white dispersion, which was subsequently transferred to an autoclave and heated at 150°C for 1 h. After cooling to room temperature, the reaction mixture was washed three times with methanol. $LaF_3$:Er,Yb nanoparticles were obtained by drying the washed product at 80°C. Finally, the synthesized nanoparticles were incubated with RB (1 mM) in dark under mild shaking condition in automated shaker for 24 hours. This nanoparticle matrix was isolated by centrifugation, washed in DI water several times until the supernatant exhibited negligible RB absorption, finally stored in phosphate-buffered saline (PBS).

Bacterial Culture plate preparation for aPDT:

The efficacy of RB-coated nanoparticles for aPDT against methicillin-resistant *Staphylococcus aureus* (MRSA) was assessed using the zone of inhibition method. A 0.1 mg/mL dispersion of the nanoparticles matrix in PBS was prepared for testing. Sterile, molten nutrient soft agar (50 mL), maintained at approximately 45°C, were combined with 500 µL of an exponentially grown MRSA suspension. This mixture was then immediately poured and evenly distributed into 3 sterile Petri dishes, allowing the soft agar to solidify. Wells with a diameter of 5 mm were aseptically created in the agar using a sterile punch. Each well was loaded with 50 µL of the nanoparticle dispersion. The wells on the plate were exposed to a 980 nm laser (1 $W/cm^2$) for varying durations of 2, 5 and 10 minutes, resulting in fluence 120, 300 and 600 $J/cm^2$, respectively. Following exposure to laser light, the culture plates were incubated at 37°C for a period of 24 h. Subsequently, the diameters of the resulting inhibition zones were precisely measured using vernier callipers to evaluate the aPDT efficacy of the nanoparticles.

**Characterizations:**

The structural, morphological, compositional and optical properties of the synthesized materials were thoroughly characterized using a range of analytical techniques. X-ray diffraction (XRD) patterns were acquired using a Rigaku diffractometer with Cu Kα radiation ($\lambda$ = 1.5406 Å) to determine the crystalline phase and structural characteristics. Particle size distribution and zeta potential in aqueous solution were measured using a Malvern Zetasizer Nano ZS90, employing a 632.8 nm laser source. Transmission electron microscopy (TEM), performed on a Philips CM200 operating at 200 keV, provided detailed information on particle size and morphology. Energy-dispersive X-ray spectroscopy (EDX), integrated with a Carl Zeiss Sigma field-emission scanning electron microscope (FE-SEM), was utilized to determine the elemental composition of the samples. Fourier-transform infrared (FTIR) spectroscopy, conducted using a Bruker ALPHA II spectrometer, was employed to identify the presence of specific functional groups. UV-visible absorption spectra were recorded using a Cintra 20 spectrophotometer (GBC Scientific Equipment Ltd.). Photoluminescence (PL) measurements were performed using a 980 nm continuous-wave (CW) diode laser as the excitation source. Spectral analysis was carried out on an Edinburgh FLS920-s fluorescence spectrometer, equipped with a double-grating monochromator. PL data were collected using a Hamamatsu R928P photomultiplier tube (PMT) detector. Thermoelectrically cooled InGaAs detectors were used for analyzing the infrared spectrum associated with singlet oxygen.

**Results and Discussion:**

The XRD pattern of $LaF_3$:Er,Yb (hereafter referred to as LF) is shown in figure-1. The observed peak positions and intensities closely match with the standard data for rhombohedral $LaF_3$ (ICDD File No. 32-0483, space group $P\overline{3}c1$), indicating the successful formation of nanoparticles in this crystal structure. [23] The sharp and well-defined peaks in the XRD pattern confirm the highly crystalline nature of the synthesized materials.

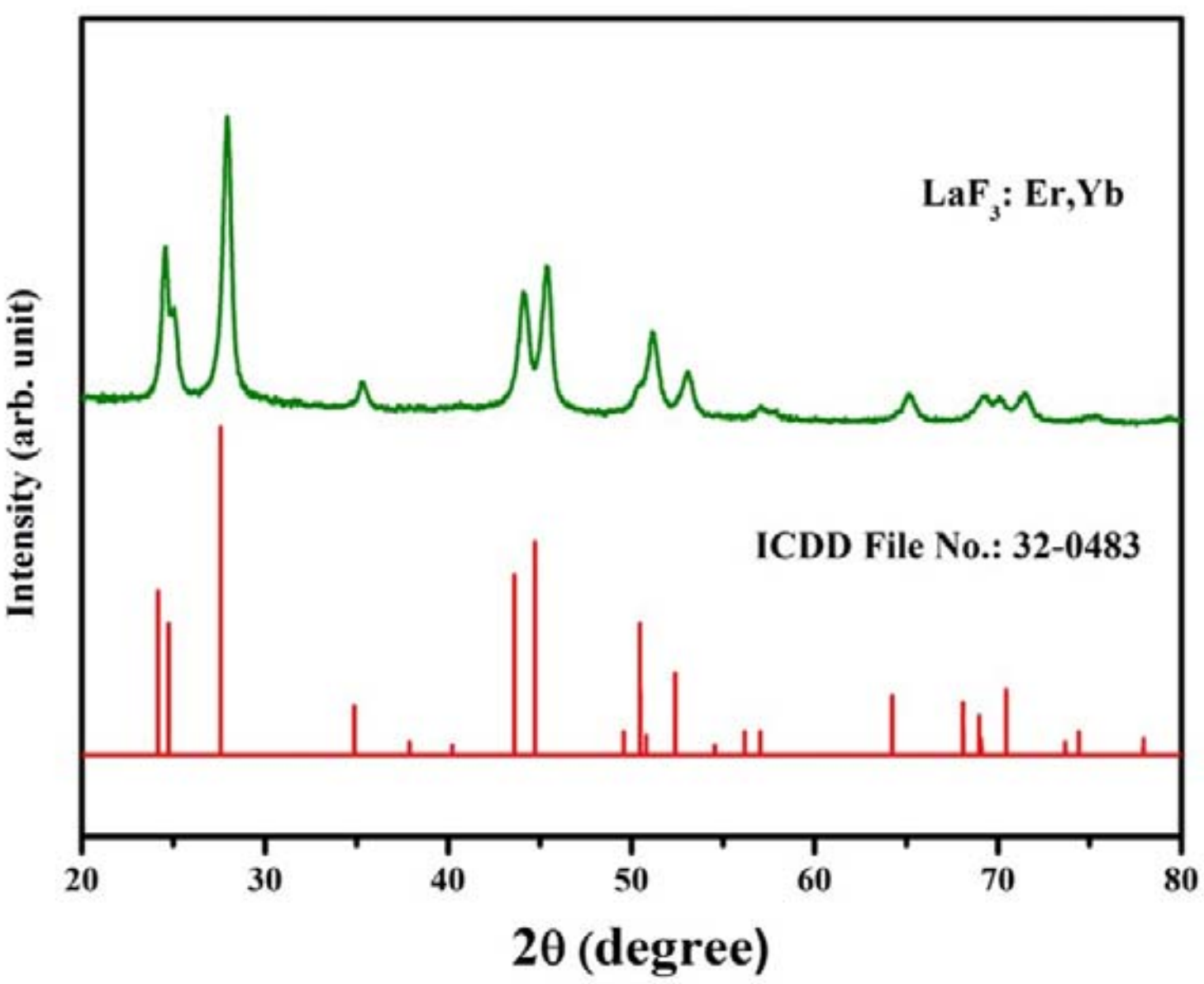

***Figure 1**: X-ray diffraction pattern of $LaF_3$:Er,Yb.*

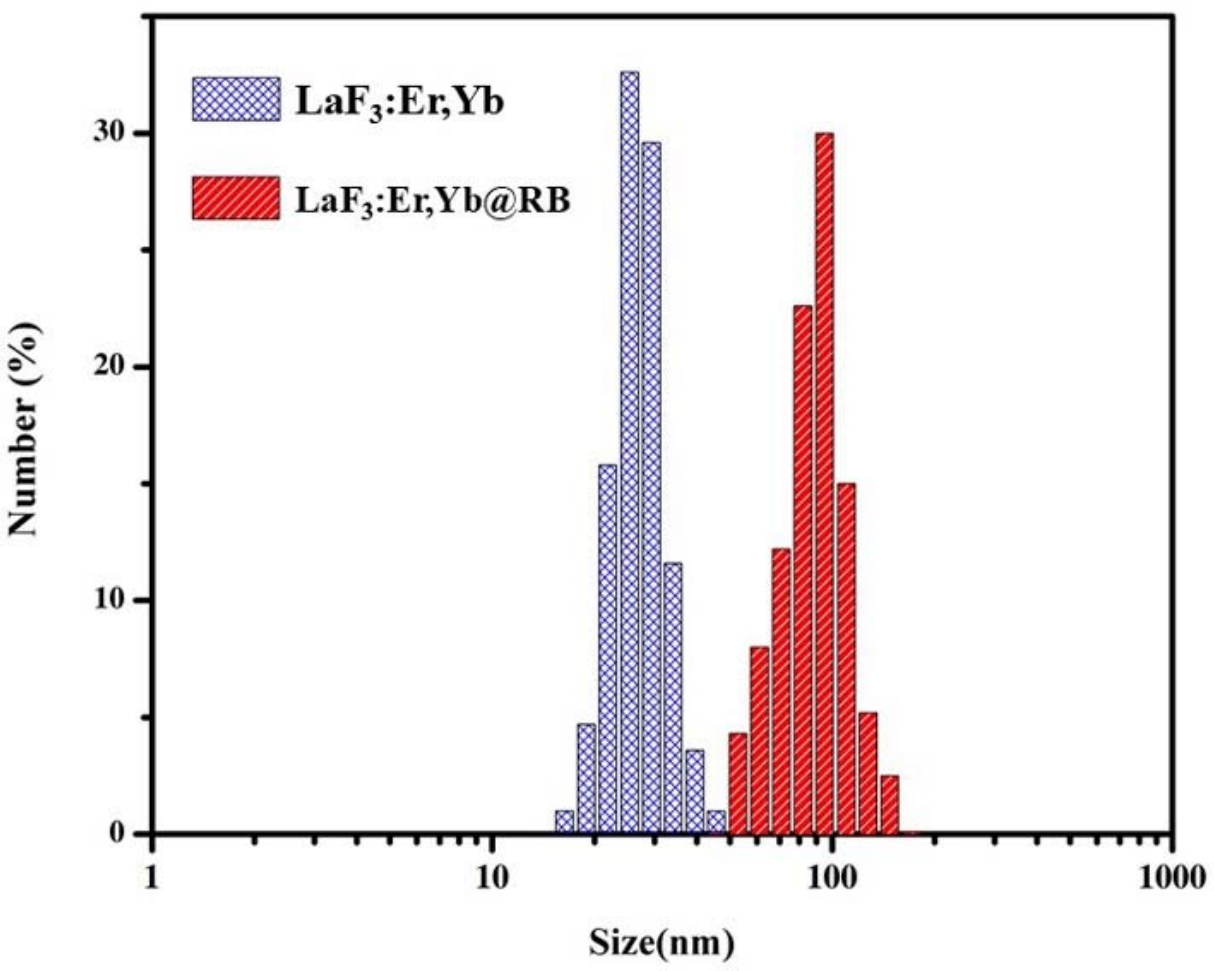

***Figure 2****: Particle size analysis using DLS, showing the distribution of bare and RB-coated* $LaF_3$*:Er,Yb nanoparticles.*

Dynamic light scattering (DLS) measurements were performed on dispersed LF and LF@RB (RB-coated LF) nanoparticles in distilled water (3 mg/mL). As shown in figure 2, the analysis revealed a narrow size distribution of approximately 20–30 nm for bare LF, indicating uniform particle synthesis. Additionally, it demonstrated a significant increase in the hydrodynamic diameter (Dh) from 24.5 ± 0.4 nm for LF to 82.3 ± 0.7 nm for LF@RB, indicating successful RB adsorption onto the nanoparticles (fig. 2). The polydispersity index (PDI) values of 0.282 ± 0.017 and 0.312 ± 0.008 for LF and LF@RB, respectively, suggest acceptable monodispersity, implying relatively uniform RB functionalization across the LF nanoparticles.[24]

Zeta potential measurements showed a positive value of +38.5 mV for bare LF nanoparticles, while LF@RB exhibited a negative value of -15.6 mV. The negative surface charge of the coated particles is likely due to the presence of Rose Bengal, an anionic dye containing sulfonate ($-SO_3^-$) and carboxyl ($-COO^-$) groups. [25,26] The shift in surface charge confirms the effective interaction between LF nanoparticles and RB molecules.

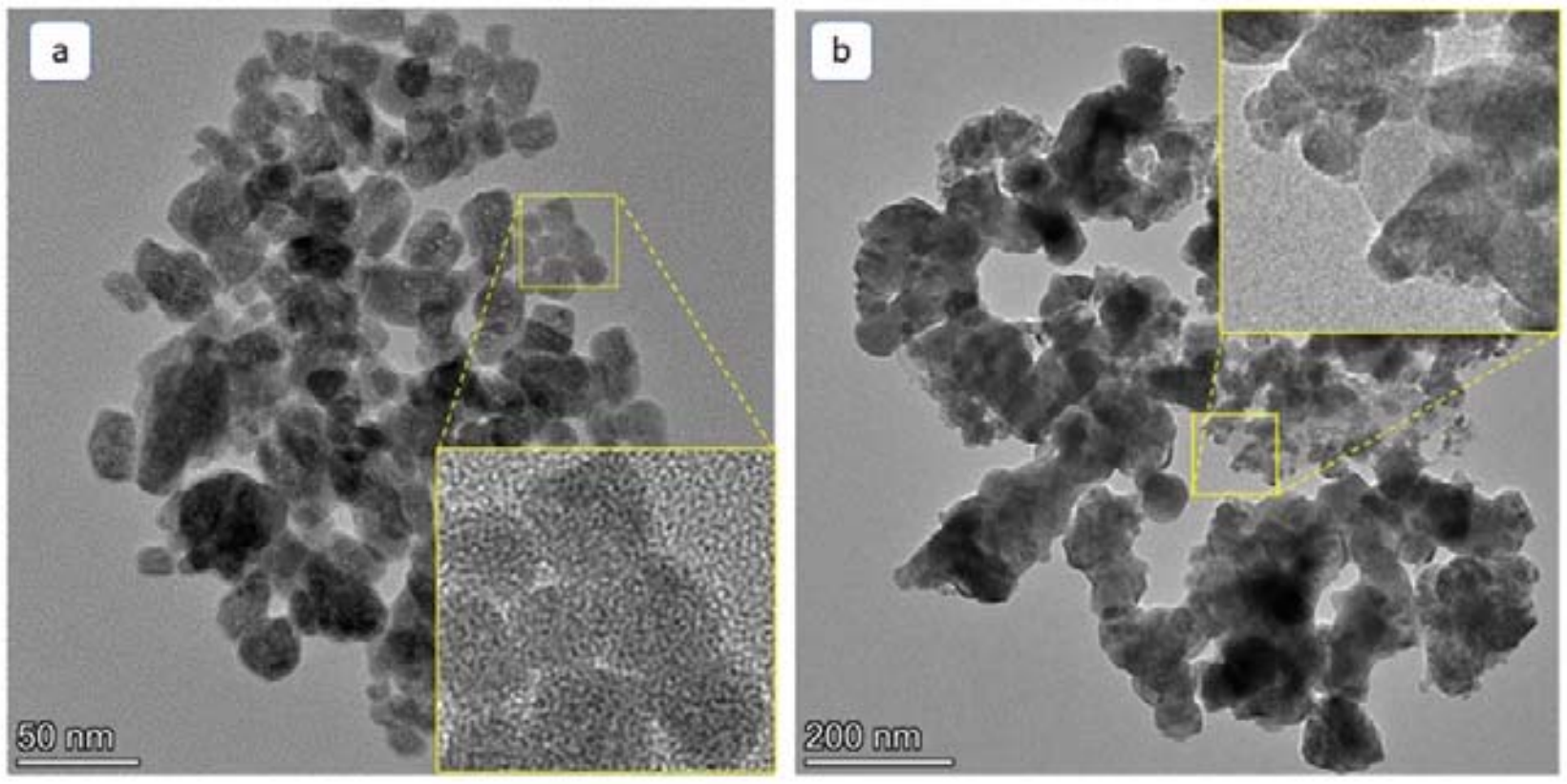

***Figure 3**: TEM images showing morphology of (a) bare $LaF_3$:Er,Yb and (b) RB-coated $LaF_3$:Er,Yb nanoparticles.*

The morphological differences between bare and RB-coated nanoparticles are highlighted in figure 3. TEM imaging showed that bare LF nanoparticles are spherical (Figure 3(a)), while coating with RB (LF@RB) resulted in aggregated particles with a distorted spherical shape (figure 3(b)). This indicates that process induced surface modifications and intermolecular forces, leading to aggregation and shape distortion.

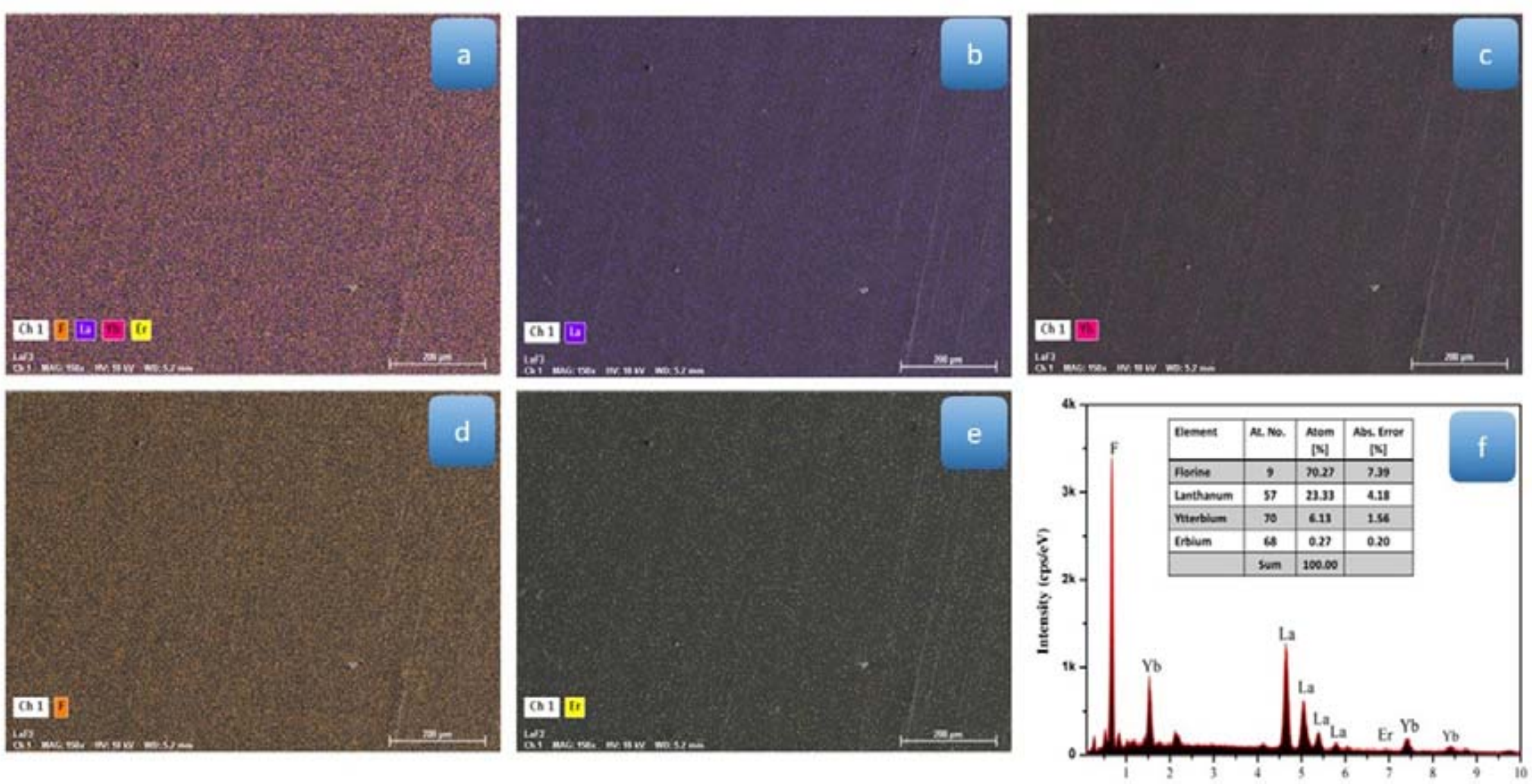

***Figure 4**: Elemental distribution in $LaF_3$:Er,Yb pellet via EDX analysis. (a) Combined elemental map. (b-e) Individual element maps: Fluorine (F), Lanthanum (La), Erbium (Er), Ytterbium (Yb). (f) EDX spectrum with inset showing relative elemental composition.*

Elemental mapping using energy dispersive X-ray analysis (EDX) was performed on a pellet (6 mm diameter) composed of LF nanoparticles, as shown in figure 4(a). The analysis revealed a uniform distribution of elements, including Lanthanum, Ytterbium, Erbium and Fluorine, as depicted in figures 4(b–e). The relative atomic percentages of these elements, determined from the EDX spectrum (figure 4(f)), provided a quantitative assessment of the synthesized nanoparticle's elemental composition.

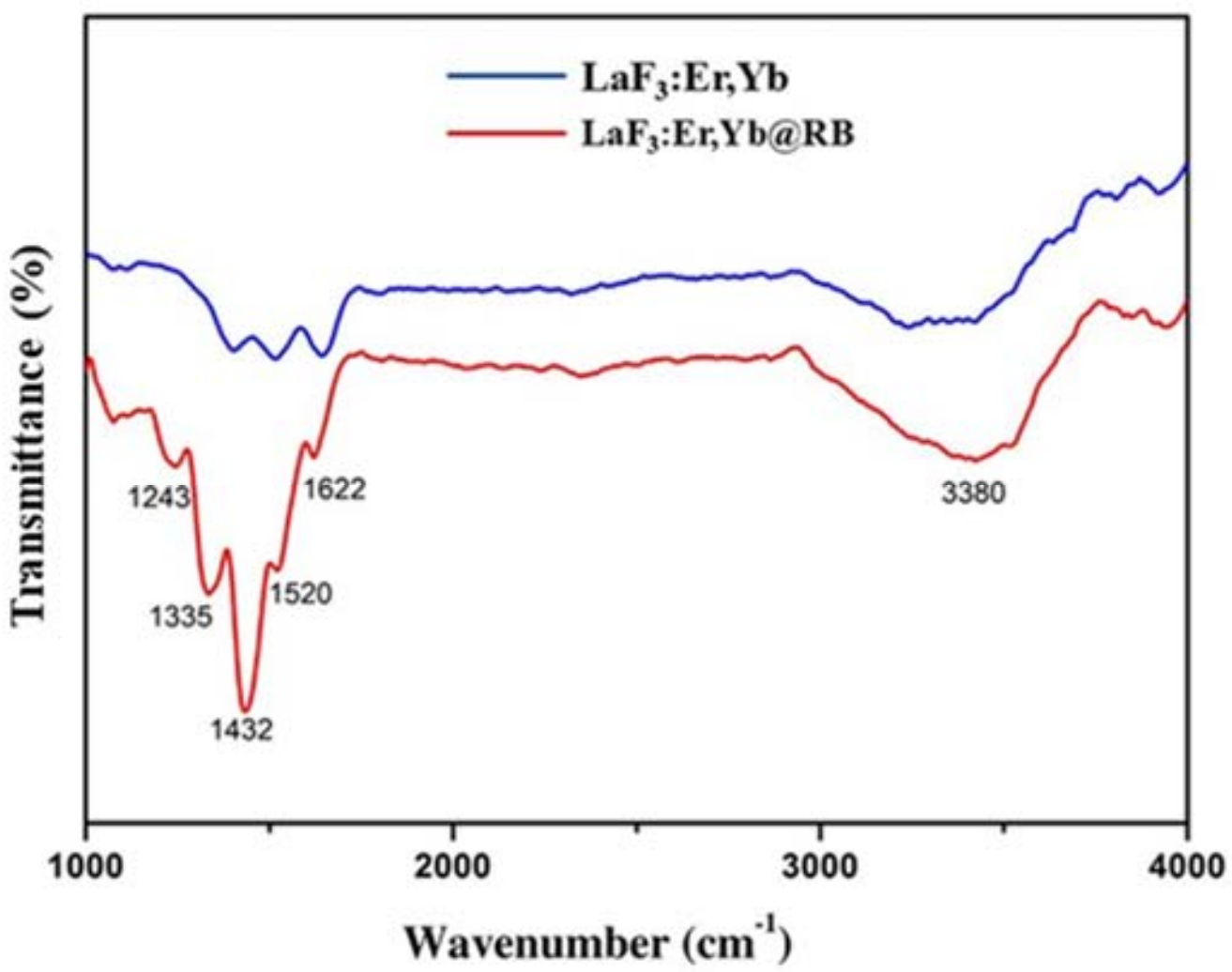

***Figure 5****: FTIR spectra of bare and RB-coated $LaF_3$:Er,Yb nanoparticles.*

FTIR spectroscopy was employed to identify the surface functional groups of the LF nanophosphor. The analysis was conducted on a powdered sample at room temperature. As shown in Figure 5, a broad band centered at 3380 $cm^{-1}$ corresponds to the O–H stretching vibration of adsorbed water.[27] The peak at 1640 $cm^{-1}$ is attributed to the stretching vibrations of the N=O group.[28] Additionally, peaks observed at 1390 $cm^{-1}$ and 1452 $cm^{-1}$ correspond to the asymmetric stretching vibrations of $NO_3^-$, indicating the potential presence of nitrate residues in the final sample.[28,29]

The successful coating of LF nanoparticles with RB was confirmed by the appearance of characteristic peaks of RB in the FTIR spectrum of LF@RB (fig. 5). Broad bands at 1432 $cm^{-1}$ and 1335 $cm^{-1}$ correspond to the N–H bending vibration and C–N stretching vibration of RB, respectively.[30] Furthermore, the spectrum of RB exhibits a carbonyl stretching (=C–O) vibration at 1622 $cm^{-1}$, along with a peak at 1520 $cm^{-1}$ associated with the aromatic (C=C) rings in the molecule.[30,31]

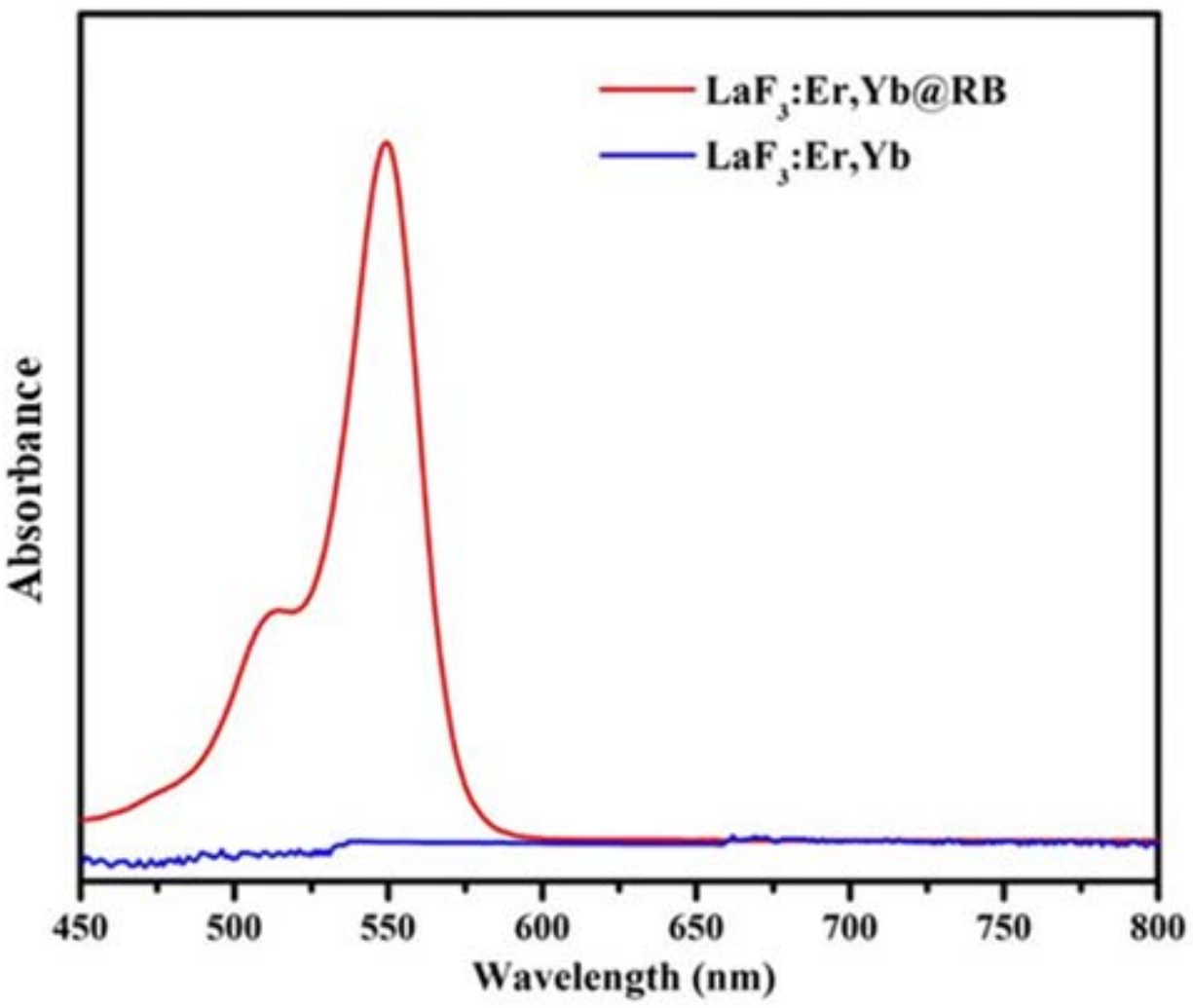

***Figure 6****: UV-Vis absorption spectra of bare and RB-coated $LaF_3$:Er,Yb nanoparticles.*

Figure 6 presents the UV-Vis absorption spectra of LF and LF@RB nanophosphors. A distinct absorption peak characteristic of RB is observed at approximately 550 nm, along with a shoulder peak at 510 nm, corresponding to the electronic excitation from the ground singlet state ($S_0$) to the first excited singlet state ($S_1$) in the RB-coated sample.[32,33] In contrast, no such absorption features are detected for the uncoated nanophosphors, confirming the successful surface coating of RB onto the nanophosphor matrix.

The upconversion luminescence properties of LF nanoparticles were characterized through emission spectral analysis, as depicted in Figure 7(a). This luminescence is a result of energy transfer processes occurring between $Yb^{3+}$ and $Er^{3+}$ ions. Incident 980 nm photons excite $Yb^{3+}$ ions from ground state ($^2F_{7/2}$) to an excited state ($^2F_{5/2}$). Due to the proximity of energy levels, excited $Yb^{3+}$ transfers energy to $Er^{3+}$ ions, acting as an activator.[34] As shown in Figure 7(a), the upconversion emission spectrum presents three primary peaks at approximately 521 nm, 548 nm and 655 nm. The green peaks (~ 521 and 548 nm) correspond to the $^2H_{11/2}$, $^4S_{3/2} \rightarrow {}^4I_{15/2}$ associated with $Er^{3+}$, while the red peak (~655 nm) originates from the $^4F_{9/2} \rightarrow {}^4I_{15/2}$ transition of the same ion.[34]A comparison of the emission intensity spectra reveals that intensity of green emission peak of LF@RB decreases compared to bare LF, which is attributed to light absorption by RB. The emission curves of bare LF and LF@RB were normalized relative to the red peak, revealing that approximately 90% of the green peak was absorbed by RB. Figure 7(b) presents the spectral overlap between the absorption peak of RB and the green emission bands of LF nanophosphors at 521 nm and 548 nm. The strong overlap between these spectral regions indicates the feasibility of Förster resonance energy transfer (FRET) from the $Er^{3+}$ activator of nanophosphor to RB.[35] This interaction is crucial for efficient energy transfer, enhancing the activation of RB and improving the generation of ROS under NIR excitation. Such a mechanism is expected to contribute significantly to the effectiveness of UCNP-mediated aPDT.

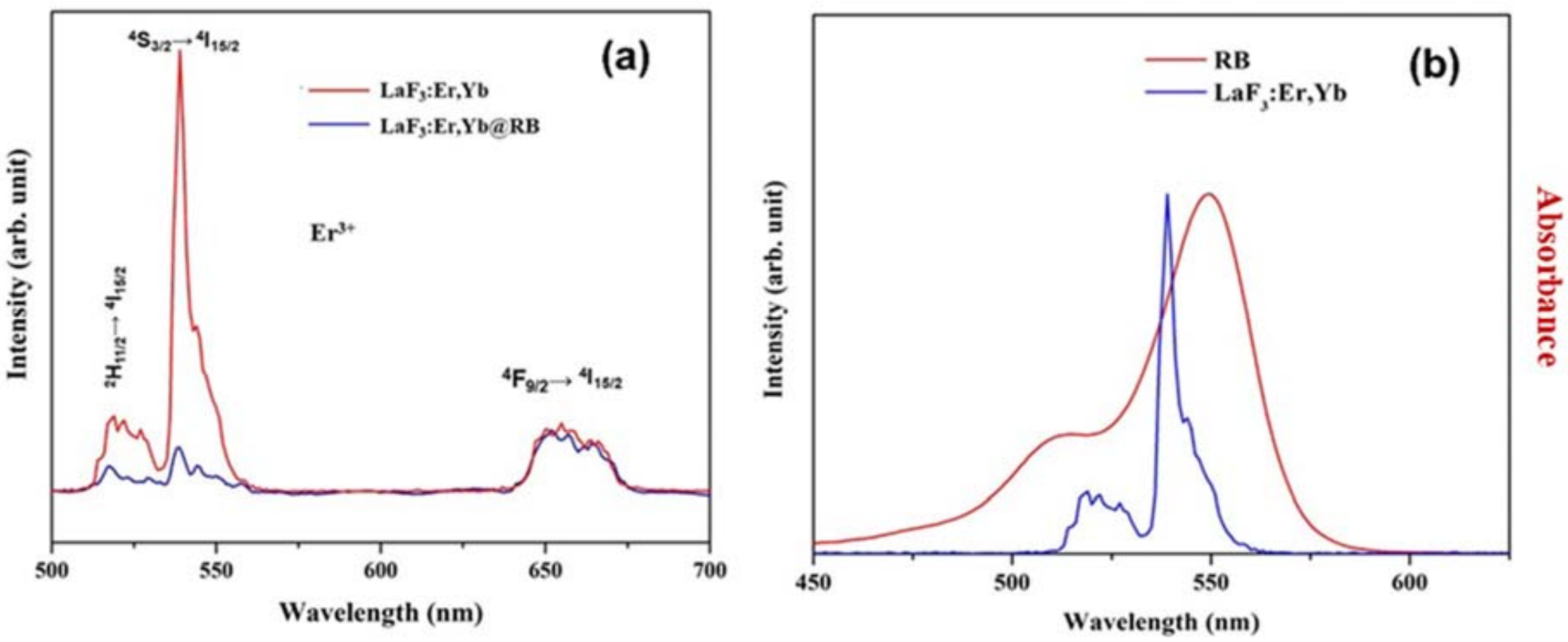

***Figure 7**: **(a)** Upconversion luminescence spectra of bare and RB-coated $LaF_3$:Er,Yb under 980 nm excitation. **(b)**: Overlap between $LaF_3$:Er,Yb emission and RB absorption.*

Assessing the generation of singlet oxygen ($^1O_2$) is pivotal in determining PDT efficacy of a PS. In this study, we evaluated the $^1O_2$ production of LF@RB nanoparticles by detecting its characteristic luminescence at approximately 1270 nm.[36,37] To enhance detection accuracy, we dispersed LF@RB nanoparticles in deuterium oxide ($D_2O$) at a concentration of 3 mg/mL, as $D_2O$ reduces the quenching effects associated with the O-H bonds in $H_2O$.[38,39] Upon excitation with a 980 nm laser (1 W/cm$^2$), UCNPs within LF@RB transferred energy to the RB coating, leading to RB excitation. This process resulted in a distinct luminescence peak centered around 1270 nm, corresponding to the transition of $^1O_2$ to its ground state, as illustrated in figure 8. The observed peak serves as a definitive spectroscopic signature, confirming the generation of singlet oxygen by the LF@RB nanoparticles under the specified experimental conditions.

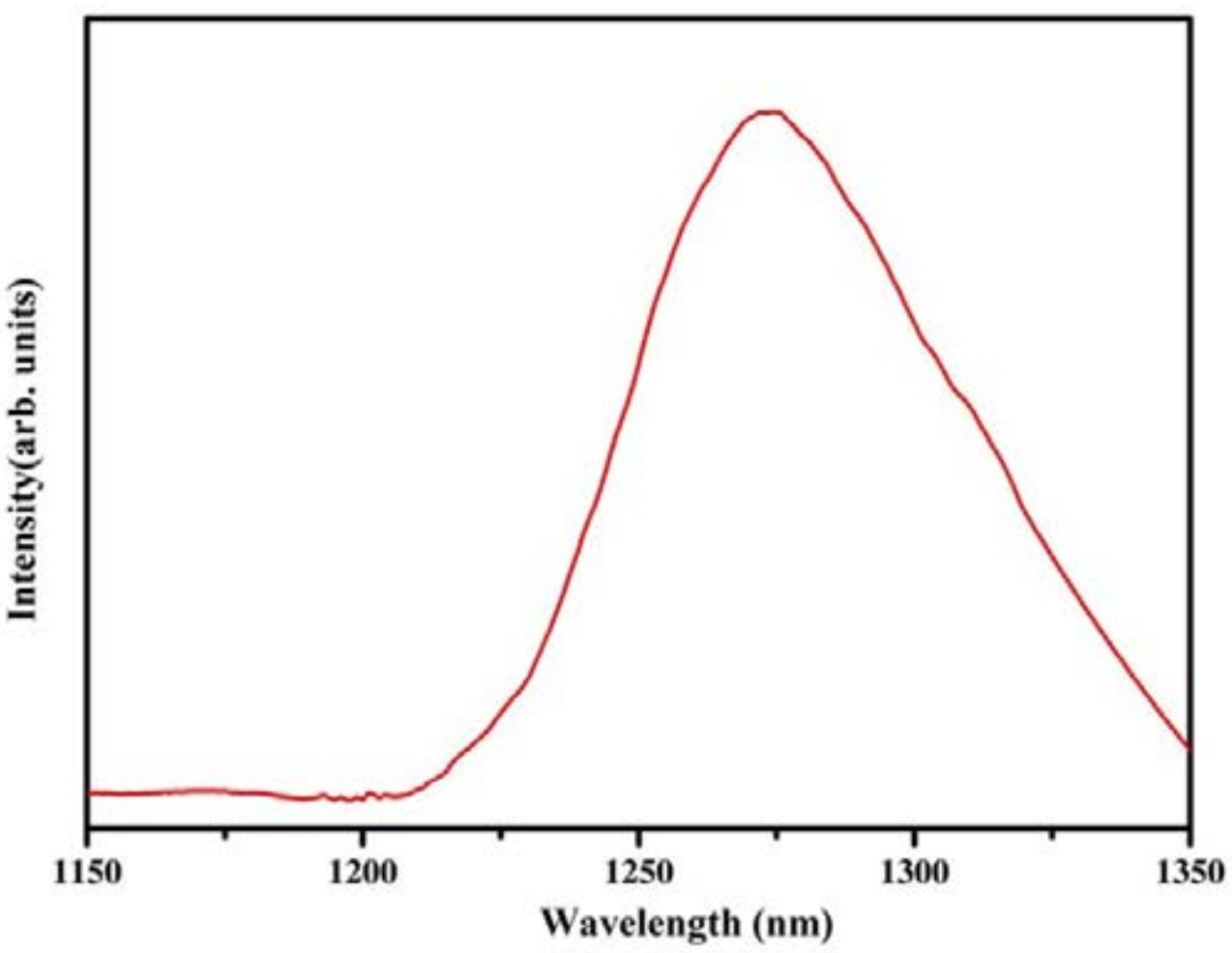

***Figure 8****: Emission Spectrum of $^1O_2$ from RB-coated $LaF_3$:Er,Yb in $D_2O$ upon 980 nm excitation.*

The photodynamic antibacterial efficacy of LF@RB was evaluated using the zone of inhibition test, as shown in figure 9(a) and 9(b). In the absence of light, LF@RB exhibited no antibacterial activity, confirming its negligible dark toxicity. However, upon irradiation with a 980 nm laser (1 W/cm$^2$) for varying durations (2, 5 and 10 minutes), distinct zones of inhibition were observed in the MRSA lawn. The diameter of the inhibition zone increased with longer irradiation times, although the difference between 5 and 10 minutes was not statistically significant, suggesting possible saturation in ROS generation within this irradiation time frame. Figure 9(b) quantitatively reflects the same trend, presenting the average inhibition zone diameters from three independent experiments, further supporting the observed time-dependent aPDT effect

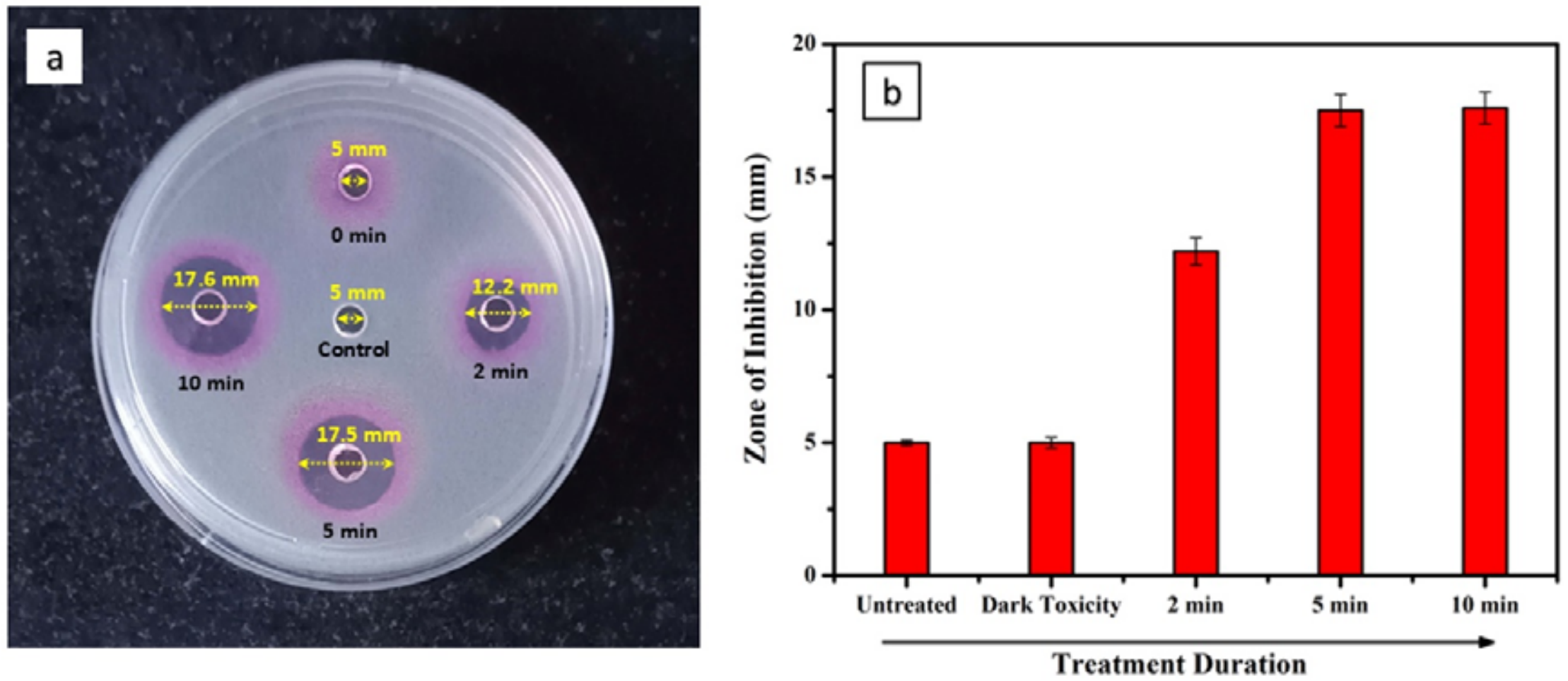

***Figure 9**: Time-dependent photodynamic antibacterial activity of RB-coated $LaF_3$:Er,Yb against MRSA. (a) Inhibition zones following 980 nm laser exposure (1 W/cm²). (b) Average inhibition zone diameters from three independent experiments.*

**Conclusions:**

This study demonstrates the synthesis and functionalization of $LaF_3:Er^{3+},Yb^{3+}$ nanoparticles with RB for NIR-activated PDT. Structural and optical characterization confirmed efficient upconversion luminescence and effective energy transfer to RB, enabling singlet oxygen generation. Zone of inhibition assay revealed significant anti-MRSA effect upon NIR irradiation, validating the therapeutic potential of these engineered nanoparticles for aPDT. The results highlight the potential of UCNP-based aPDT for deeper tissue treatments, overcoming the limitations of traditional visible-light based RB-aPDT. Future work will prioritize nanoparticle optimization, *in vivo* biocompatibility assessments and clinical translation pathways to harness this technology for practical medical applications.

**Acknowledgments:**

The authors wish to thank Dr. Himanshu Srivastava of *Accelerator Physics and Synchrotrons Utilization Division, RRCAT, Indore, India* for his contribution in performing TEM imaging of the samples.

**Author Contributions:**

**Pratik Deshmukh**: Conceptualization, Methodology, Writing – Original Draft. **Nandini Ahuja**: Visualization, Investigation. **Bhumika Sharma**: Data Curation. **Khageswar Sahu**: Validation. **Srinibas Satapathy**: Conceptualization and Supervision. **Shovan Kumar Majumder**: Funding acquisition and supervision.

## Declarations:

**Conflict of interest:** On behalf of all authors, the corresponding author states that there is no conflict of interest.

**Funding:** No external funding was received for this study.

**Data and code availability:**

Data and code will be made available from corresponding authors upon reasonable request.

**Supplementary information:** Not Applicable.

**Ethical approval**: Not Applicable.

**FIGURE CAPTIONS:**

***Figure 1****: X-ray diffraction pattern of $LaF_3$:Er,Yb.*

***Figure 2****: Particle size analysis using DLS, showing the distribution of bare and RB-coated $LaF_3$:Er,Yb nanoparticles.*

***Figure 3****: TEM images showing morphology of (a) bare $LaF_3$:Er,Yb and (b) RB-coated $LaF_3$:Er,Yb nanoparticles.*

***Figure 4****: Elemental distribution in $LaF_3$:Er,Yb pellet via EDX analysis. (a) Combined elemental map. (b-e) Individual element maps: Fluorine (F), Lanthanum (La), Erbium (Er), Ytterbium (Yb). (f) EDX spectrum with inset showing relative elemental composition.*

***Figure 5****: FTIR spectra of bare and RB-coated $LaF_3$:Er,Yb nanoparticles.*

***Figure 6****: UV-Vis absorption spectra of bare and RB-coated $LaF_3$:Er,Yb nanoparticles.*

***Figure 7****:* ***(a)*** *Upconversion luminescence spectra of bare and RB-coated $LaF_3$:Er,Yb under 980 nm excitation.* ***(b)****: Overlap between $LaF_3$:Er,Yb emission and RB absorption.*

***Figure 8****: Emission Spectrum of $^1O_2$ from RB-coated $LaF_3$:Er,Yb in $D_2O$ upon 980 nm excitation.*

***Figure 9**: Time-dependent photodynamic antibacterial activity of RB-coated $LaF_3$:Er,Yb against MRSA. (a) Inhibition zones following 980 nm laser exposure (1 W/cm²). (b) Average inhibition zone diameters from three independent experiments.*